# Econophysics of Macro-Finance: Local Multi-fluid Models and Surface-like Waves of Financial Variables

Victor Olkhov

TVEL, Kashirskoe sh. 49, Moscow, 115409, Russia,

victor.olkhov@gmail.com

Abstract

This paper models macro financial variables alike to financial fluids with local interactions and describes surface-like waves of Investment and Profits. We regard macro-finance as ensemble of economic agents and use their risk ratings as coordinates on economic space. Aggregations of agent's financial variables with risk coordinates $x$ on economic space define macro-financial variables as function of $x$. We describe evolution and interactions between macro-financial variables alike to financial fluids by hydrodynamic-like equations. Minimum and maximum risk grades define most secure and most risky agents respectively. That determines borders of macro-finance domain that is filled by economic agents. Perturbations of agent's risk coordinates near risk borders of macro domain cause disturbances of macro financial variables like Investment and Profits. Such disturbances can generate waves that propagate along risk borders. These waves may exponentially amplify perturbations inside of macro domain and impact financial sustainability. We study simple model Investment and Profits and describe linear approximation of steady state distributions of Investment and Profits on macro-finance domain that fulfill "dreams" of Investors: "more risks – more Profits". We describe Investment and Profits waves on risk border of economic space *alike to* surface waves in fluids. We present simple examples that specify waves as possible origin of time fluctuations of macro-financial variables. Description of possible steady state distributions of macro financial variables and financial risk waves on economic space could help for better policy-making and managing sustainable macro-finance.

JEL: C00, E00, F00, G00 [1]

---

[1] This research did not receive any specific grant or financial support from TVEL or funding agencies in the public, commercial, or not-for-profit sectors and was performed on my own account only.

# 1. Introduction

Macro-finance modeling plays crucial role for financial policy and management. Most intrinsic macro-finance relations concerns influence of economic and financial fluctuations on asset pricing [1-4]. This paper studies one possible origin of financial fluctuations. Complexity of macro-financial relations causes diversity of financial fluctuations nature and modeling. We suppose that origin of most economic and fluctuations is determined by propagation of waves induced by perturbations of financial variables. This paper describes macro financial waves that may propagate along risk borders of macro financial domain on economic space alike to surface waves in fluids. Financial waves require certain space where such waves can propagate. We use economic space as ground for macro-finance model. We model macro-finance as ensemble of agents each described by numerous financial and economic variables. Economic space allows describe evolution of economic agents and their financial variables like Profits and Investment, Assets and Liabilities and describe propagation of macro financial waves [5-9]. Risk ratings of economic agents are treated as their coordinates and that allow describe agent's financial variables as functions on economic space. Financial variables of agents at point $x$ on economic space defines macro financial variables as functions of coordinates $x$ on economic space. For example, correct sum (without doubling) of agent's Profits at point $x$ defines macro Profits as function of $x$ on economic space. Changes of agent's financial variables like Investment or Profits define changes of macro financial variables – Investment or Profits. Financial transactions between economic agents evaluate changes of agent's variables. For example Investment of agent *A* into Assets of agent *B* rise or fall due to rise or fall of Investment transactions from agent *A* to agent *B*. Simplest assumption on dynamics of financial transactions that allows develop quantitative macro-finance model states that transactions occur between agents with same coordinates on economic space. Let call such approximation as *local* macro-finance model. In *local* model we describe interaction between macro financial variables by linear differential operators. As we show for *local* macro-finance approximation disturbances of macro variables can induce financial waves that have parallels to acoustic waves in fluids [9]. These waves can propagate on economic space and their amplitudes may amplify in time by exponent. Wave generation, propagation and interactions are the most general and important properties that impact dynamics of any complex system and macro-finance is definitely such a case. Development of macro-finance models requires description of possible macro financial wave processes.

In present paper we study one more type of waves that may exist in macro-finance and describe simple examples of waves actions on macro financial variables like Investment or Profits. Up now risk ratings of huge corporations and banks are provided by international rating agencies [10-12]. Let propose that risk assessment methodologies permit estimate risk ratings for all economic agents and risk ratings take continuous values. Let regard ratings of single risk as coordinates of one-dimensional space and simultaneous rating assessments of *n* risks as measurements of agent's coordinates on *n*-dimensional economic space. Minimum risk ratings correspond to most secure and maximum ratings correspond to most risky agents. Thus all macro-finance agents fill domain between minimum and maximum risk ratings. Thus macro-finance under the action of two risks can be presented as ensemble of agents that fill rectangular domain on $R^2$. Economic agents with minimum and maximum risk ratings define borders of this macro-finance rectangle. Financial and economic perturbations and random nature of risk may induce disturbances of agent's risk ratings and their variables near macro borders. That can bring disturbances of macro financial variables. Such disturbances on borders of macro-finance domain have certain parallels to disturbances on surface of fluids those induce surface waves [13,14]. In this paper for *local* macro-finance model we describe evolution and interactions between macro variables by hydrodynamic-like equations and describe Investment surface-like waves induced by perturbations on macro-finance risk border. Such simple waves on macro borders are *alike to* surface waves in fluids. We say "*alike to*" to underline vital diversities between nature of macro-finance and nature of physical fluids. Nevertheless certain *parallels* between them allow develop quantitative macro-finance model and derive wave equations on macro financial variables. Correct description of any complex systems like macro-finance requires usage of adequate complex methods and models. Complexity of hydrodynamic-like macro-finance model reflects internal complexity of financial interrelations and helps better describe real evolution of financial system.

Our approach to macro-finance as a system of economic agents on economic space is completely different from general equilibrium models [15], economic decision-making [16], behavioral economics [17], agent-based economics [18], spatial economics [19,20] and etc. We don't argue any equilibrium state and replace question: "Why economic agents take certain decisions?" with a different one: "How agent's financial variables describe macro financial variables?" We hope that our approach to macro-finance might be interesting for readers.

The rest of the paper is organized as follows. In the Section 2 we give problem setup. In Section 3 we explain meaning of hydrodynamic-like approximation of macro-finance. In Section 4 for simple model of interaction between macro Investment and Profits we derive wave equations of disturbances of Profits due to perturbations of risk coordinates near risk border of macro-finance and study simple wave solutions. Conclusions are in Section 5.

## 2. Model Setup and Economic Space

For reader's convenience we give below brief definition of economic space, introduce macro financial variables and explain usage of hydrodynamic-like equations for macro-finance modeling according to [5-9].

*2.1. Economic space*

All agents of macro-finance have numerous financial variables as Profits and Investment, Assets and Debts, Capital and Credits and etc. Agent's variables determine corresponding macro financial variables. Thus description of evolution of agent's variables help model evolution of macro financial variables. Let propose that it is possible estimate risk ratings of all macro-finance agents. If so let use their risk ratings as their coordinates on certain economic space. That allows attribute agent's financial variables by same coordinate. For example agent's Profits can be described by agent's coordinates. That allows define macro financial variables as functions of coordinate on economic space. Indeed, sum (without doubling) of Profits of all agents with coordinates $x$ define macro financial Profits as function of $x$. Sum or integral of Profits by $x$ over economic space define common notion of macro-finance Profits as function of time only. Description of macro financial variables as functions on economic space allows model interactions between macro variables by tools of mathematical physics. We propose that such complication of macro-finance modeling methods correspond to internal complexity of macro-finance processes. Adequate description of such complex systems as macro-finance requires complex methods. Introduction of economic space and transition from usual description of macro financial variables as functions of time to financial variables treated as functions of time and *coordinates* on economic space enhance modeling tools and methods.

International rating agencies [10-12] estimate risk ratings of huge corporations and banks and these ratings are widespread in finance. Risk ratings take values of risk grades and noted as *AAA, BB, C* and so on. Let treat risk grades like *AAA, BB, C* as points $x_1, x_2,.. x_m$ of discreet space. Let propose, that risk assessments methodologies can be extended to estimate risk ratings for all macro-finance agents. Many risks impact macro-finance. Let regard grades of single risk as points of one-dimensional space and simultaneous risk assessments of $n$ different risks as measurements of agent's coordinates on $n$-dimensional space. Let propose, that risk assessments methodologies can be generalized in such a way that risk grades can take continuous values and define space $R$. Thus risk grades of $n$ different risks establish $R^n$.

Let define economic space as any mathematical space that map economic agents by their risk ratings as space coordinates. Let take that number $n$ of risks ratings measured simultaneously defines dimension *n* of economic space. Let put positive direction along each risk axis as risk growth direction. Let assume that econometrics provide sufficient info about risk ratings and financial variables of each agent. These assumptions require significant development of current econometrics and statistics. We believe that accuracy of current U.S. National Income and Product Accounts system [12] give confidence that all these problems could be solved.

There are a lot of different risks that impact macro-finance. To develop reasonable macro-finance model one should select few major risks and neglect minor risks. Selection of *n* risks with major impact on macro-finance agents defines economic space $R^n$ that determine initial state of macro-finance. Assessment and comparison of different risks and their influence on macro-finance establish tough problems and such models should be developed. Risk assessments methodologies and procedures, comparison of risk influence on performance of macro-finance can establish procedures alike to measurement theory in physics and improve macro-finance modeling and forecasting.

Financial risks have random nature and can unexpectedly arise and then vanish. Thus some current risks that define initial *n*-risks representation of economic space $R^n$ can accidentally disappear and new major risks may come to play. To forecast macro-finance in a time term *T* one should predict *m* main risks that may play major role in a particular time term and define $R^m$ representation of economic space. Such a set of *m* risks determine target state of economic space $R^m$. Transition of macro-finance modeling from initial economic space $R^n$ to a target economic space $R^m$ requires description of decline of action of initial set of *n* risks and growth of new *m* risks.

Current models of macro-finance describe relations between macro variables and their fluctuations as Profits and Investment, Credits and Debts, each treated as function of time. Introduction of economic space gives ground for description macro-finance variables as functions of time and *coordinates*. That gives a new look on macro-finance modeling.

*2.2. Macro financial variables*

In previous sub-section we defined macro financial variables at point *x* on economic space as correct sum (without doubling) of corresponding values of same variables of agents with coordinate *x*. However random motion of agents on economic space and stochastic variations of their financial variables makes above definition also random. To introduce regular macro financial variables let take following considerations.

To introduce regular macro financial variables as functions of time and coordinates on economic space let average randomness and fluctuations associated with granularity of separate economic agents and their financial variables. Random changes of agent's risk coordinates make ensemble of agents *looks like "financial gas"*. For instance agent's Profits change randomly with random change of agent's risk ratings. Averaging of financial variables by probability distributions allow neglect agent's granularity and describe macro financial variables *alike to "continuous finance media"* or by hydrodynamic-like approximation.

Let explain way for transition from description of financial variables of separate agents to description of macro financial variables as functions of time and coordinates on economic space [5,9]. For brevity let call economic agents as economic particles or e-particles and economic space as e-space. Let define macro financial variables as sum of corresponding values of agent's variables with coordinates $x$ on e-space. Let assume that e-particles on e-space are "*independent*" so, for example, sum of Profits of any $k$ e-particles equal total Profits of group of these $k$ e-particles. Aggregation of Profits of e-particles at point $x$ on e-space define Profits as function of time $t$ and $x$. Integral of Profits by $dx$ over e-space equals Profits of entire macro-finance as function of time $t$. E-particles and their financial variables are under random motion on e-space. Hence sum of Profits of e-particles near point $x$ also has random values. To obtain regular financial variable like regular Profits at point $x$ let average sum of Profits near point $x$ by probability distribution $f$. Let state that distribution $f$ define probability to observe $N(x)$ e-particles with Profits equal $a_1,...a_{N(x)}$. That determine Profits macro financial density at point $x$ on e-space (Eq.(2.1) below). Due to motion of e-particles on e-space such macro financial density behaves *alike to "financial fluid"* – it can flow. To describe motion of such Profit's fluid let define velocity of such a fluid. Let mention that velocities of e-particles are not additive variables and their sum doesn't define velocity of Profits motion. To define velocities of Profits fluid correctly one should define "Profit's impulses" at point $x$ as product of Profits $a_j$ of e-particle $j$ and it's velocity $\boldsymbol{v_j}$ (Eq. (2.2) below). Such "Profit's impulses" $a_j \boldsymbol{v_j}$ - are additive variables and sum of "Profit's impulses" can be averaged by similar probability distribution $f$. Densities of Profits and densities of Profits impulses define velocities of Profits financial fluid (Eq.(2.3) below). Let mention that different financial fluids can have different velocities. For example flow of Investment on e-space can have velocity higher then flow of Profits, nevertheless they are determined by motion of same e-particles. Let present above considerations in a more formal way.

Let assume that there are $N(x)$ e-particles at point $x$. Let state that velocities of e-particles at point $x$ equal $v=(v_1,... v_{N(x)})$ and each e-particle on e-space $R^n$ at moment $t$ is described by $l$ extensive financial variables $(u_1,...u_l)$. Extensive financial variables are additive and admit averaging by probability distributions. Intensive financial variables, like Prices or Interest Rates, are not additive and cannot be averaged directly. Let assume that values of financial variables equal $u=(u_{1i},...u_{li})$, $i=1,..N(x)$. Each extensive financial variable $u_j$ at point $x$ defines macro financial variable $U_j$ as sum of variables $u_{ji}$ of $N(x)$ e-particles at point $x$

$$U_j = \sum_i u_{ji} ; \quad j = 1,..l; \quad i = 1,...N(x)$$

To describe motion of financial variable $U_j$ let establish additive variable alike to impulse in physics. For e-particle $i$ let define financial impulses $p_{ji}$ as product of extensive variable $u_j$ that takes value $u_{ji}$ and its velocity $v_i$:

$$p_{ji} = u_{ji} v_i \tag{1.1}$$

Thus if e-particle has $l$ extensive financial variables $(u_1,...u_l)$ and velocity $v$ then it has $l$ impulses $(p_1,p_2,..p_l)=(u_1v,...u_lv)$. Let define impulse $P_j$ of financial variable $U_j$ as

$$\boldsymbol{P}_j = \sum_i u_{ji} \boldsymbol{v}_i ; \quad j = 1,..l; \quad i = 1,...N(x) \tag{1.2}$$

Let introduce distribution function $f=f(t,x;U_1,..U_l, \boldsymbol{P}_1,..\boldsymbol{P}_l)$ that determine probability to observe financial variables $U_j$ and impulses $\boldsymbol{P}_j$ at point $x$ at time $t$. $U_j$ and $\boldsymbol{P}_j$ are determined by corresponding values of e-particles that have coordinates $x$ at time $t$. They take random values at point $x$ due to random motion of e-particles on e-space. Averaging of $U_j$ and $\boldsymbol{P}_j$ by probability distribution function $f$ allows establish transition from approximation that takes into account financial variables of separate e-particles to continuous "*financial media*" or hydrodynamic-like approximation of macro-finance that neglect e-particles granularity and describe macro financial variables as regular functions of time and coordinates on e-space. Let define financial density functions $U_j(t,x)$ as

$$U_j(t,\boldsymbol{x}) = \int U_j \, f(t,\boldsymbol{x},U_1,...U_l,\boldsymbol{P}_1,..\boldsymbol{P}_l) \, dU_1..dU_l d\boldsymbol{P}_1..d\boldsymbol{P}_l \tag{2.1}$$

and impulse density functions $\boldsymbol{P}_j(t,x)$ as

$$\boldsymbol{P}_j(t,\boldsymbol{x}) = \int \boldsymbol{P}_j \, f(t,\boldsymbol{x},U_1,...U_l,P_1,..P_l) \, dU_1..dU_l d\boldsymbol{P}_1..d\boldsymbol{P}_l \tag{2.2}$$

That allows define e-space velocities $v_j(t,x)$ of financial densities $U_j(t,x)$ as

$$U_j(t,\boldsymbol{x}) \boldsymbol{v}_j(t,\boldsymbol{x}) = \boldsymbol{P}_j(t,\boldsymbol{x}) \tag{2.3}$$

Densities $U_j(t,x)$ and impulses $\boldsymbol{P}_j(t,x)$ are determined as mean values of aggregates of corresponding financial variables of separate e-particles with coordinates $x$. Functions $U_j(t,x)$ can describe macro financial densities of Profits and Investment, Loans and Debts and so on.

## 3. Hydrodynamic-like macro-finance approximation

In previous section we introduced notions of macro financial densities, impulses and velocities. We use *language, wording* of hydrodynamics to outline parallels to macro-finance modeling. Our transition from description of financial variables of separate e-particles (economic agents) to description of continuous financial densities and their velocities is *similar* *to* transition from kinetics as description of multi-particle system to continuous media or hydrodynamic approximation in physics. We repeat that nature of macro-finance has nothing common with multi-particle systems in physics. Nevertheless transition from description of multi-agent system to hydrodynamic-like approximation gives new look on macro-finance modeling.

Definitions of macro financial densities and velocities as functions of time and coordinates on e-space permit describe their evolution by hydrodynamic-like equations. Let follow [14] and for any extensive financial variable *U* and it's velocity ***v*** on e-space let take Continuous Equations (3.1) and Equations of Motion (3.2) in form [5,9]:

$$\frac{\partial U}{\partial t} + div(\boldsymbol{v}U) = Q_1 \tag{3.1}$$

$$U\left[\frac{\partial \boldsymbol{v}}{\partial t} + (\boldsymbol{v}\cdot\nabla)\boldsymbol{v}\right] = \boldsymbol{Q}_2 \tag{3.2}$$

Financial meaning of Eq.(3.1, 3.2) is very simple. Left side of Eq.(3.1) describes change of *U(t,**x**)* in unit volume on e-space $R^n$ at point ***x***. It can change due to time derivative $\partial U/\partial t$ and due to *div(U**v**)* that equals flux of *U**v*** through surface of unit volume at point ***x***. $Q_1$ describe any external factors that can change *U(t,**x**)*. Left side of Equations of Motion describes same variation of impulse ***P**(t,**x**) = U(t,**x**)**v**(t,**x**)*. Taking into account Continuity Equations we simplified left side of Equations of Motion and take it as Eq.(3.2.). $\boldsymbol{Q}_2$ describe any factors that change left side *(3.2)*.

Let state that factors $Q_1$ and $\boldsymbol{Q}_2$ for Eq.(3.1, 3.2) depend on densities $U_j(t,\boldsymbol{x})$, $\boldsymbol{v}_j(t,\boldsymbol{x})$ that are *different* from *U(t,**x**)*, ***v**(t,**x**)*. Let call that variables $U_j(t,\boldsymbol{x})$, $\boldsymbol{v}_j(t,\boldsymbol{x})$ are *conjugate* to variables *U(t,**x**)*, ***v**(t,**x**)* if variables $U_j(t,\boldsymbol{x})$, $\boldsymbol{v}_j(t,\boldsymbol{x})$ determine $Q_1$ and $\boldsymbol{Q}_2$ factors in right hand side of hydrodynamic-like Eq.(3.1, 3.2). For example, Profits may have *conjugate* variables like Investment, Cost of Capital, Sales and their velocities. Let state, that *conjugate* variables or *conjugate* financial fluids define right hand side factors $Q_1$ and $\boldsymbol{Q}_2$ in Continuity Equation and Equation of Motion. As we mentioned in Introduction, for simplification this paper studies *local* financial transactions between e-particles on e-space. Here we take into account transactions between e-particles with nearly same coordinates only. Such simplification is *alike to* modeling collisions between e-particles and allow describe factors $Q_1$ and $\boldsymbol{Q}_2$ by linear differential operators on *conjugate* densities and their velocities. We use this assumption in the next Section. *Non-local* macro-finance model that takes into account financial "action-at-a-distance" between agents on e-space is presented in [9].

Eq.(3.1, 3.2) present at least two ways for macro-finance modeling. First one allows describe evolution of *U* and it's velocity ***v*** under action of determined external factors $Q_1$ and $\boldsymbol{Q}_2$. It can be used in the assumption that evolution of *U* does not impact on $Q_1$ and $\boldsymbol{Q}_2$. This approximation describes *U* in fixed macro-finance environment.

Second approach takes into account mutual dependence between *U* and *conjugate* variables that determine factors $Q_1$ and $Q_2$. If two, tree or more macroeconomic variables define mutual dependence of factors $Q_1$ and $Q_2$ we obtain a system of two, three or more Eq. like (3.1, 3.2). For example, if equations on $U_1$ are determined by factors $Q_1$ and $Q_2$ that depends on one *conjugate* variable $U_2$ and vise versa then we obtain self-consistent equations on two macro financial variables in a closed form.

The simplest model that presents Eq.(3.1, 3.2) in a closed form describes interaction between two *conjugate* e-fluids. For that case let study derivation of risk wave equations on e-space. Up now terms "waves" are used in economics and macro-finance to describe Kondratieff waves, inflation waves, crisis waves, etc. All these issues describe time oscillations of variables only. Description of waves requires a space where such waves can propagate. Introduction of e-space gives ground for macro-finance waves studies.

## 4. Investment-Profits interaction model and waves

Let study simple macro-finance model of interaction between two *self-conjugate* variables as Investment and Profits on *2*-dimensional e-space $R^2$ with coordinates (*x,y*), so all e-particles are under action of two risks presented by axes *X* and *Y*. Risk ratings of e-particles define their coordinates on e-space $R^2$. Let state that risk ratings are reduced by minimum risk grade that correspond to most secure e-particles and maximum risk grade that refers to most risky e-particles. For simplicity let assume that coordinates (*x,y*) of e-particles on e-space $R^2$ follows *0 ≤ x ≤ X* for risk axis *X* and *0 ≤ y ≤ Y* for risk axis *Y*. Coordinates (0,0) describe most trusted e-particles and (*X,Y*) reflect most risky e-particles. Thus e-particles on e-space $R^2$ fill rectangle with sides *(0,X)* and *(0,Y)* along axes *X* and *Y*. All macro-finance variables are determined on macro rectangle *(0,X;0,Y)*. Four boundaries of rectangle are free surfaces of macro domain on e-space $R^2$ that determine macro-finance variables of most secure and most risky e-particles. Economic or financial processes may disturb risk coordinates of e-particles at macro boundary and that can cause disturbances of macro financial variables. These perturbations of macro variables on free surface of macro domain may dissipate, may grow up or may generate waves that can propagate along macro-finance surface or inside macro domain. Let call such waves of macro financial variables as surface-like risk waves. Description of surface-like waves of Demand and Supply, Investment and Profits, Income and Consumptions are very important for macro-finance modeling as they show new way of impact of financial shocks and perturbations on macro financial sustainability due to propagation of surface waves over e-space. Wave generation, propagation and interactions describe core properties for almost all complex physical systems. We believe that such complex system as macro-finance should admit variety wave processes that describe interrelations between macro financial variables and govern macro-finance sustainability. Up now there are no econometric evidence of such macro financial surface waves but we propose that such processes can exist and can be observed. Let study surface-like risk waves for simple models.

Let study simple model of possible interaction between two *self-conjugate* macro financial variables as Investment and Profits. Investment density *I(t,x)* describes Investment into e-particles with coordinates *x* and density *P(t,x)* describes Profits generated by Investment in point *x*. Due to above consideration let describe such a model by hydrodynamic-like Eq.(3.1, 3,2) for Investment *I(t,x)* and it' velocity *v* and for Profits *P(t,x)* and it's velocity *u*. Let study interactions between Investment and Profits on macro rectangle with sides *(0, X)* and *(0, Y)* along *X* and *Y* axes. Let assume that in steady state macro-finance domain has boundary that is determined by relations *y=Y*. Let assume that in steady state densities of Investment *I* and Profits *P* equal zero for *y>Y*. Let study waves that can be generated by perturbations of macro densities and their velocities near boundary *y=Y*. Let define perturbations of boundary surface as *y= ξ(t,x)*. Interactions between Investment and Profits require that macro boundary *y= ξ(t,x)* should be common for both. Otherwise interaction between Investment and Profits will be broken. Time derivation of *y= ξ(t,x)* determines *y*-velocities $v_y$ and $u_y$ of both densities on macro surface *y= ξ(t,x)* as:

$$\frac{\partial}{\partial t}\xi(t,x) = v_y(t,x,y=\xi(t,x)) = u_y(t,x,y=\xi(t,x)) \qquad (4.1)$$

To derive wave equations on disturbances of Investment and Profits let follow derivation of surface wave equations in physical fluids [13,14]. Let assume that velocities *v* and *u* of e-fluids are determined by potentials *φ* and *ψ* as

$$\boldsymbol{v} = \nabla\varphi \quad ; \quad \boldsymbol{u} = \nabla\psi \qquad (4.2)$$

Let neglect non-linear terms in Equations of Motion and take them as:

$$I_0 \frac{\partial \boldsymbol{v}}{\partial t} = Q_{21} \quad ; \quad P_0 \frac{\partial \boldsymbol{u}}{\partial t} = Q_{22} \qquad (4.3)$$

where $I_0$ and $P_0$ – Investment and Profits constants that we define below. Let account dependence of Investment *I(t,x,y)* and Profits *P(t,x,y)* on time and coordinates in Continuity Equations (3.1) as

$$\frac{\partial I}{\partial t} + \nabla \cdot (I\boldsymbol{v}) = Q_{11} \quad ; \quad \frac{\partial P}{\partial t} + \nabla \cdot (P\boldsymbol{u}) = Q_{12} \qquad (4.4)$$

To go further let define interactions between Investment and Profits and determine factors $Q_1$ and $\boldsymbol{Q}_2$. As we mentioned above we study *local* interaction between e-particles on e-space and propose that e-particles at point *x* carry transactions with e-particles near same point. Such assumptions simplify real interrelations between macro financial variables. Nevertheless even simplified models demonstrate extreme complexity and diversity of internal financial processes on e-space. Assumption on *local transactions* between e-particles means that interrelations between macro densities and their velocities are also *local*. Let describe such interactions by simple linear differential operators. Let put $Q_{11}$ for Continuity equation (4.4) on Investment be proportional to Profits velocity *y*-component $u_y$. Let put $Q_{12}$ for Continuity Equation (4.4) on Profits be proportional to Investment velocity *y*-component $v_y$. Indeed, positive Profits velocity $u_y$ increase Investment flow *Iv* and Investment are attracted by flow of Profits. Let take Continuity Equation on Investment as:

$$\frac{\partial I}{\partial t} + \nabla \cdot (I\boldsymbol{v}) = Q_{11} = a_1 u_y \ ; \ a_1 > 0 \tag{5.1.1}$$

As well positive Investment velocity $v_y$ may decrease Profits flow *Pu* because additional Investment flow increase mutual competition between e-particles and decrease available Profits. Let take Continuity Equations (4.4) on Profits as:

$$\frac{\partial P}{\partial t} + \nabla \cdot (P\boldsymbol{u}) = Q_{12} = a_2 v_y \ ; \ a_2 < 0 \tag{5.1.2}$$

To define $Q_{2i}$ factors for Equations of Motion (4.3) let assume that acceleration of Investment velocity $\boldsymbol{v}$ is proportional to gradient of Profits *P*. Thus Investment flows in the direction of higher Profits and

$$Q_{21} \sim b\nabla P \ ; \ b > 0 \tag{5.2.1}$$

Let assume that acceleration of Profits velocity $\boldsymbol{u}$ is proportional to gradient of Investment *I*. As well area with higher Investment may reduce velocity of Profits flow and take $Q_{22}$ as:

$$Q_{22} \sim d\nabla I \ ; \ d < 0 \tag{5.2.2}$$

Let introduce *"financial accelerations"* **h**=($h_x,h_y$) and **g**=($g_x, g_y$) that act on Investment and Profits respectively along risk axes *X* and *Y* and prevent agents from excess Investment into risky area. Such *"financial accelerations"* **h** and **g** may describe collective fears of investors, managers and shareholders to loose their Investment and total corporate Value due to excess risks. As well taking Investment risks reflect common financial "dream": "more risks - more Profits". As we show below *"financial accelerations"* describe simple model of steady state distribution of Profits and Investment with linear dependence on risks *X* and *Y*. Let introduce *"financial accelerations"* **h**=($h_x,h_y$) acting on Investment and **g**=($g_x, g_y$) acting on Profits along axes *X* and *Y* by potentials *H* and *G* as follows:

$$\frac{d}{dx}G = g_x \ ; \ \frac{d}{dy}G = g_y \ ; \ \frac{d}{dx}H = h_x \ ; \ \frac{d}{dy}H = h_y \tag{5.2.3}$$

and write Equations of Motion as:

$$I_0 \frac{\partial}{\partial t} v_x = Q_{21x} = -g_x P_0 + b \frac{\partial}{\partial x} P \quad ; \quad I_0 \frac{\partial}{\partial t} v_y = Q_{21y} = -g_y P_0 + b \frac{\partial}{\partial y} P \quad (5.3.1)$$

$$P_0 \frac{\partial}{\partial t} u_x = Q_{22x} = -h_x I_0 + d \frac{\partial}{\partial x} I \quad ; \quad P_0 \frac{\partial}{\partial t} u_y = Q_{22y} = -h_y I_0 + d \frac{\partial}{\partial y} I \quad (5.3.2)$$

Relations (4.2) allow present Eq.(5.3; 5.4) as

$$I_0 \frac{\partial}{\partial t}\frac{\partial}{\partial x} \varphi = -g_x P_0 + b \frac{\partial}{\partial x} P \quad ; \quad I_0 \frac{\partial}{\partial t}\frac{\partial}{\partial y} \varphi = -g_y P_0 + b \frac{\partial}{\partial y} P$$

$$P_0 \frac{\partial}{\partial t}\frac{\partial}{\partial x} \psi = -h_x I_0 + d \frac{\partial}{\partial x} I \quad ; \quad P_0 \frac{\partial}{\partial t}\frac{\partial}{\partial y} \psi = -h_y I_0 + d \frac{\partial}{\partial y} I$$

Then Investment $I$ and Profits $P$ take form

$$I(t,x,y) = I_0\left(1 + \frac{1}{d} H(x,y)\right) + \frac{P_0}{d} \frac{\partial}{\partial t} \psi(t,x,y) \quad (5.4.1)$$

$$P(t,x,y) = P_0\left(1 + \frac{1}{b} G(x,y)\right) + \frac{I_0}{b} \frac{\partial}{\partial t} \varphi(t,x,y) \quad (5.4.2)$$

It is obvious that potentials $H$ and $G$ Eq.(5.2.3) describe model of steady state distributions of Investment and Profits determined by Eq.(5.4.1;5.4.2) on e-space for case $\varphi=\psi=0$. We use "steady state" notion to underline distinctions with equilibrium states in the meaning of statistical physics. We affirm no similarities between equilibrium notions in statistical physics and steady distributions in macro-finance and economics. We believe that macroeconomics and macro-finance have no "equilibrium" states in the meaning of statistical physics. We consider macro-finance as strongly "non-equilibrium" system which evolution can be modeled by transitions from one steady state to another. Description of steady state distributions of macro financial variables on e-space establish important and tough problem. Description of macro-finance steady states could allow study dependence of steady distributions of macro financial variables on risk coordinates and model financial policy that could manage transitions from one macro financial steady state to another. Modeling possible macro financial steady states on e-space requires further studies. In this paper for simplicity we regard (5.2.3) with potentials $H$ and $G$ as linear functions with **h** and **g**-*constant*.

$$H(x,y) = h_x x + h_y y \quad ; \quad G(x,y) = g_x x + g_y y \quad (5.4.3)$$

Eq.(5.4.1-3) present Investment $I(t,x,y)$ and Profits $P(t,x,y)$ as:

$$I(t,x,y) = I_0\left(1 + \frac{1}{d}[h_x(x-X) + h_y(y-Y)]\right) + \frac{P_0}{d} \frac{\partial}{\partial t} \psi(t,x,y) \quad (5.5)$$

$$P(t,x,y) = P_0\left(1 + \frac{1}{b}[g_x(x-X) + g_y(y-Y)]\right) + \frac{I_0}{b} \frac{\partial}{\partial t} \varphi(t,x,y) \quad (5.6)$$

$I_0$ and $P_0$ are values of Investment and Profits in a steady state at most risky point of macro domain on e-space with coordinates $(X,Y)$

$$I(t,X,Y) = I_0 \quad ; \quad P(t,X,Y) = P_0$$

Due to Eq.(5.2.1;5.2.2) $b>0$ and $d<0$ hence Investment $I(t,X,Y)$ at most risky point $(X,Y)$ on macro domain rectangle $[0,X;0,Y]$ take minimum value $I_0$. As well Profits $P(t,X.Y)$ at most risky point *(X,Y)* on macro domain take maximum value $P_0$. Due to Eq.(5.2.2; 5.5) Investment *I(t,x,y)* decrease from maximum value at most riskless and secure point with coordinates *(0,0)* to minimum value at most risky point *(X,Y)*. Meanwhile due to Eq.(5.2.1; 5.6) Profits *P(t,x,y)* have maximum value at most risky point *(X,Y)* and minimum value at most secure point *(0,0)*. Thus Eq.(5.5;5.6) give simple example of steady state distribution of Investment and Profits over macro-finance domain on economic space and describe common relationships – "risks increase Profits". These simplified linear relations means that Investment and Profits in a steady state at most secure point with coordinates *(0,0)* equal:

$$I(0,0) = I_0\left[1 - \frac{1}{d}(h_x X + h_y Y)\right]; \ P(0,0) = P_0\left[1 - \frac{1}{b}(g_x X + g_y Y)\right], b > g_x X + g_y Y$$

Let assume that small perturbations $y = \xi(t,x)$ of border *y=Y* don't change values of Investment and Profits on border *y= Y*. In a steady state for *y= Y*

$$I(t,x,Y) = I_0[1 + \frac{h_x}{d}(x - X)] \ ; \ P(t,x,Y) = P_0[1 + \frac{g_x}{b}(x - X)]$$

Thus on surface $y= \xi(t,x)$ we obtain

$$I_0\left[1 + \frac{h_x}{d}(x - X) + \frac{h_y}{d}(\xi(t,x) - Y)\right] + \frac{P_0}{d}\frac{\partial}{\partial t}\psi(t,x,y = \xi(t,x)) = I_0[1 + \frac{p_x}{d}(x - X)]$$

$$P_0\left[1 + \frac{g_x}{b}(x - X) + \frac{g_y}{b}(\xi(t,x) - Y)\right] + \frac{I_0}{b}\frac{\partial}{\partial t}\varphi(t,x,y = \xi(t,x)) = P_0[1 + \frac{g_x}{b}(x - X)]$$

Hence obtain:

$$\xi(t,x) = Y - \frac{P_0}{h_y I_0}\frac{\partial}{\partial t}\psi(t,x,y = \xi(t,x)) = -\frac{I_0}{g_y P_0}\frac{\partial}{\partial t}\varphi(t,x,y = \xi(t,x)) \quad (5.7)$$

Eq.(5.7) determine relations between $h_y$ and $g_y$

$$I_0^2 h_y = P_0^2 g_y$$

Eq.(4.1; 5.7) give:

$$\frac{\partial}{\partial t}\xi(t,x) = \frac{\partial}{\partial y}\psi = \frac{\partial}{\partial y}\varphi = -\frac{I_0}{g_y P_0}\frac{\partial^2}{\partial t^2}\varphi(t,x,y = \xi(t,x)) \quad (5.8)$$

Let use Continuity Equations (5.1.1; 5.1.2) to obtain equations on potentials $\varphi$ and $\psi$. Let substitute Eq.(5.5; 5.6) into Continuity Equations (5.1.1; 5.1.2) and neglect non-linear terms with potentials and "*financial accelerations*" **h**=($h_x,h_y$) and **g**=($g_x, g_y$). Then equations on potentials $\varphi$ and $\psi$ take form:

$$\left(P_0\frac{\partial^2}{\partial t^2} - a_1 d\frac{\partial}{\partial y}\right)\psi(t,x,y) = -dI_0\Delta\varphi \ ; \ \left(I_0\frac{\partial^2}{\partial t^2} - a_2 b\frac{\partial}{\partial y}\right)\varphi(t,x,y) = -bP_0\Delta\psi \quad (6.1)$$

$$\Delta = \frac{\partial^2}{\partial x^2} + \frac{\partial^2}{\partial y^2}$$

Treatment of Eq.(6.1) admits two approximations. The first and simplest approximation is based on assumption that divergences of velocities **v** and **u** equal zero and thus

$$\nabla \cdot v = \nabla \cdot u = 0 \implies \Delta\varphi = \Delta\psi = 0 \quad (6.2)$$

If so Eq.(6.1) on potentials $\varphi$ and $\psi$ take form:

$$\left(P_0 \frac{\partial^2}{\partial t^2} - a_1 d \frac{\partial}{\partial y}\right)\psi(t,x,y) = 0 \; ; \; \left(I_0 \frac{\partial^2}{\partial t^2} - a_2 b \frac{\partial}{\partial y}\right)\varphi(t,x,y) = 0 \tag{6.3}$$

The second approximation don't use Eq.(6.2) and hence Eq.(6.1) on potentials take form:

$$\left[\left(P_0 \frac{\partial^2}{\partial t^2} - a_1 d \frac{\partial}{\partial y}\right)\left(I_0 \frac{\partial^2}{\partial t^2} - a_2 b \frac{\partial}{\partial y}\right) - bdP_0 I_0 \Delta^2\right]\varphi = 0 \tag{6.4}$$

Let take potential as

$$\varphi = \psi = \cos(kx - \omega t) f(y - Y) \; ; \; f(0) = 1 \tag{6.5}$$

Eq.(5.8) for potential (6.5) gives relations on function $f(y)$:

$$\frac{\partial}{\partial y} f(0) = \frac{\omega^2 I_0}{g_y P_0} > 0 \tag{6.6}$$

Let study both cases described by Eq.(6.2;6.3) and Eq.(6.4) with potentials (6.5;6.6).

## 4.1. "Non-compressible" financial fluids and surface-like waves

The first approximation is based on Eq.(6.2; 6.3) and describe case with zero divergence of velocities. In hydrodynamics fluids with zero divergence of velocity are called as "non-compressible" fluids and we use the same notion for this approximation. For potentials (6.5) Eq.(6.3) gives equations on function $f(y)$ and define constraints on $I_0$ and $P_0$:

$$\left(P_0 \omega^2 + a_1 d \frac{\partial}{\partial y}\right) f(y) = 0 \; ; \; \left(I_0 \omega^2 + a_2 b \frac{\partial}{\partial y}\right) f(y) = 0 \tag{7.1}$$

$$f(y) = \exp\left(-\frac{P_0 \omega^2}{a_1 d}(y - Y)\right) \; ; \; I_0 = \frac{a_2 b}{a_1 d} P_0 \tag{7.2}$$

Relations (5.1.1-5.2.2; 5.8; 6.6) give $a_1 > 0; a_2 < 0 \; ; \; b > 0; d < 0$

$$-\frac{P_0}{a_1 d} = \frac{I_0}{g_y P_0} > 0 \; ; \; -\frac{g_y}{a_1 d} P_0^2 = I_0 \tag{7.3}$$

Due to Eq.(7.2) and Eq. (5.2.1; 5.2.2)

$$P_0 = -\frac{a_2 b}{g_y} > 0 \; ; \; I_0 = -\frac{(a_2 b)^2}{a_1 d g_y} > 0 \; ; \; \frac{P_0 \omega^2}{a_1 d} < 0 \tag{7.4}$$

and function $f(y)$ as (7.2) dissipate inside macro domain for $y < Y$. Thus "non-compressible" relations (6.2) define values (7.4) of Investment $I_0$ and Profits $P_0$ at point $(X,Y)$ and hence define functions (5.5; 5.6). "Non-compressible" Eq.(6.2) determine relations between frequency $\omega$ and wave number $k$. Due to Eq.(6.2; 6.5;6.6)

$$\left(\frac{P_0 \omega^2}{a_1 d}\right)^2 = k^2 \; ; \; k = \pm \frac{a_2 b}{a_1 d} \frac{\omega^2}{g_y} \; ; \; \omega = \sqrt{\frac{a_1 d}{a_2 b} g_y k} \tag{7.5}$$

Wave group velocity $c$ equals

$$c = \frac{d\omega}{dk} = \sqrt{\frac{a_1 d}{4 a_2 b} \frac{g_y}{k}} = \sqrt{\frac{a_1 d}{8\pi a_2 b} g_y l} \tag{7.6}$$

Here $l=2\pi/k$ is a wavelength. Thus short waves with small wavelength $l<<X$ propagate along border $y=Y$ of macro domain slower that long waves $l\approx X$. In other words, long-scale disturbances on risk border $y=Y$ propagate along risk border faster then short-scale disturbance. Macro border perturbations defined by function $y=\xi(t,x)$ by Eq.(4.1; 5.7) give:

$$\xi(t,x) = Y - \frac{P_0}{p_y I_0}\omega \sin(kx - \omega t) = Y - \frac{a_2 b}{a_1 d}\frac{\omega}{g_y}\sin(kx - \omega t)$$

Now let study second case that describe surface-like waves for "compressible" financial fluids model due to Eq.(6.4).

## 4.2. "Compressible" financial fluids and surface-like waves

The second case presents surface-like waves for "compressible" financial fluids described by Eq.(6.4). Let substitute potentials as (6.5) into Eq.(6.4) and obtain equation of function f(y):

$$[\left(P_0\omega^2 + a_1 d \frac{\partial}{\partial y}\right)\left(I_0\omega^2 + a_2 b \frac{\partial}{\partial y}\right) - bdP_0 I_0 (k^4 - 2k^2 \frac{\partial^2}{\partial y^2} + \frac{\partial^4}{\partial y^4})]f(y) = 0 \quad (8.1)$$

Eq.(8.1) is ordinary differential equation of forth order:

$$\left(q_4 \frac{\partial^4}{\partial y^4} + q_3 \frac{\partial^3}{\partial y^3} + q_2 \frac{\partial^2}{\partial y^2} + q_1 \frac{\partial}{\partial y} + q_0\right)f(y) = 0 \quad (8.2)$$

$$q_4 = -bdP_0 I_0 \;;\; q_3 = 0 \;;\; q_2 = a_1 d\, a_2 b + 2k^2 bdP_0 I_0$$

$$q_1 = \omega^2(P_0 a_2 b + I_0 a_1 d) \;;\; q_0 = I_0 P_0 \omega^4 - bP_0 d I_0 k^4$$

Due to Eq.(5.1.1; 5.1.2; 5.2.1; 5.2.2) $q_2$ may be positive or negative and

$$q_4 > 0 \;;\; q_1 < 0 \;;\; q_0 > 0 \quad (8.2.1)$$

Eq.(8.2) has four roots $s_1,...s_4$ and their sum

$$\sum_{i=1,4} s_i = -\frac{q_3}{q_4} = 0$$

Roots $s_i$ $i=1,..4$ are determined by coefficients $q_i$, $i=0,1,..4$ and these relations between roots $s_i$ and coefficients $q_i$ define constraints between frequency $\omega$, wave number $k$ and $I_0$, $P_0$ and other parameters. These constraints define dependence between frequency $\omega$ and wave number $k$ and thus determine wave group velocity alike to Eq.(7.5; 7.6). Characteristic polynomial of Eq.(8.2):

$$q_4 s^4 + q_3 s^3 + q_2 s^2 + q_1 s + q_0 = 0 \quad (8.3)$$

Each root $s$ of Eq.(8.3) define partial solution of Eq.(8.1; 8.2) as

$$f(y - Y; s) = \exp(s(y - Y)) \quad (8.4)$$

Solutions $f(y)$ can be linear combination of (8.4) defined by roots of (8.3). Eq.(8.1) may has tree types of solutions due to roots of characteristic polynomial (8.3) [22]:

1. All roots $s_1,...s_4$ of Eq.(8.3) are real. Due to Vieta theorem and Eq.(8.2.1) two roots should be positive and two roots should be negative. Solutions $f(y)$ of Eq.(8.1) may have form

$$f(y - Y) = \sum_i \lambda_i \exp(s_i(y - Y)) \;;\; \sum_i \lambda_i = 1 \;;\; \sum_i \lambda_i s_i = \frac{\omega^2 I_0}{g_y P_0} > 0 \quad (8.5)$$

2. Two roots $s_{1,2}$ of Eq.(8.3) are real and two roots $s_{3,4}=r+/-i\theta$ are complex conjugate. Due to Vieta theorem and Eq.(8.2.1) if real part $r>0$ then $s_{1,2}<0$ and vice versa. Real solutions $f(y)$ of Eq.(8.1) may have form:

$$f(y-Y) = \sum_{i=1,2} \lambda_i \exp(s_i(y-Y)) + \exp(r(y-Y))[\lambda_3 \cos(\theta(y-Y)) + \lambda_4 \sin(\theta(y-Y))]$$

$$\sum_{i=1,3} \lambda_i = 1 \; ; \; \sum_{i=1,2} \lambda_i s_i + \lambda_3 r + \lambda_4 \theta = \frac{\omega^2 I_0}{g_y P_0} > 0 \tag{8.6}$$

3. Four roots of Eq.(8.3) are complex $s_{1-4} =+/- r+/-i\theta_{1,3}$

$$f(y-Y) = \exp(r(y-Y))[\lambda_1 \cos(\theta_1(y-Y)) + \lambda_3 \sin(\theta_1(y-Y))]$$
$$+ \exp(-r(y-Y))[\lambda_2 \cos(\theta_3(y-Y)) + \lambda_4 \sin(\theta_3(y-Y))]$$

$$\lambda_1 + \lambda_2 = 1 \; ; (\lambda_1 - \lambda_2)r + \lambda_3 \theta_1 + \lambda_4 \theta_3 = \frac{\omega^2 I_0}{g_y P_0} > 0 \tag{8.7}$$

Simplest solution of Eq.(8.1) for real root $s>0$ of Eq.(8.3) gives potentials $\varphi$ and $\psi$ as:

$$\varphi = \psi = \cos(kx - \omega t) \exp(s(y-Y)); \; f(y-Y) = \exp(s(y-Y)); \; \frac{\omega^2 I_0}{g_y P_0} = s > 0 \tag{9.1}$$

Function $y=\xi(t,x)$ due to Eq.(5.9; 9.1) takes form:

$$\xi(t,x) = Y - \frac{\omega}{g_y} \sin(kx - \omega t) = Y - \sqrt{\frac{sP_0}{g_y I_0}} \sin(kx - \omega t) \tag{9.2}$$

Investment $I$ and Profits $P$ densities waves near stationary surface $y=Y$ take form

$$I(t,x,Y) = I_0[1 + \frac{h_x}{d}(x-X)] + \frac{P_0 \omega}{d} \sin(kx - \omega t) \tag{9.3}$$

$$P(t,x,Y) = P_0[1 + \frac{g_x}{b}(x-X)] + \frac{I_0 \omega}{b} \sin(kx - \omega t) \tag{9.4}$$

Integral of Investment density $I(t,x,Y)$ (9.3) by $dx$ along border $y=Y$ by over $(0,X)$ on e-space $R^2$ define Investment at $y=Y$ as function of time:

$$I(t,Y) = I_0[X - \frac{h_x X^2}{2d}] + 2\frac{P_0 \omega}{dk} \sin\left(\omega t - k\frac{X}{2}\right) \sin\left(\frac{X}{2}k\right) \tag{9.5}$$

Eq.(9.5) show that Investment and Profits surface-like waves (9.2-9.4) can induce time fluctuations of Investment and Profits of entire macro-finance near steady state risk border $y=Y$ with frequency $\omega$. Thus irregular time fluctuations of Investment or Profits near maximum risk rating $Y$ may indicate action of random surface-like waves that propagate on border $y=Y$ along another risk axis $X$. Investment and Profits flows near steady states $(x_0, y_0)$ are alike to liquid flows that are induced by surface waves in hydrodynamics [14]. Unit volumes of Investment and Profits flows circulate near risk surface $y=Y$ by circle trajectories:

$$\frac{dx}{dt} = v_x = -k \sin(kx_0 - \omega t) \exp(s(y_0 - Y)) \; ; \; \frac{dy}{dt} = v_y = s \cos(kx_0 - \omega t) \exp(s(y_0 - Y))$$

Due to small scale of circulations near steady states $(x_0, y_0)$

$$x - x_0 = -\frac{k}{\omega} \cos(kx_0 - \omega t) \exp(s(y_0 - Y)) \; ; \; y - y_0 = -\frac{s}{\omega} \sin(kx_0 - \omega t) \exp(s(y_0 - Y))$$

Trajectories of Investment and Profits flows and of e-particles that constitute these flows:

$$(x - x_0)^2 + (y - y_0)^2 = \frac{k^2}{sg_y}[1 + \frac{s^2-k^2}{k^2} sin^2(kx_0 - \omega t)] exp(2s(y_0 - Y))$$

For simplest solution Eq.(9.1) $s>0$ and hence scale of circulations and disturbances of Investment and Profits exponentially decrease inside macro financial domain $y<Y$.

Now let show that solutions (8.5-8.7) of Eq.(8.1) may describe *exponential amplification* of small disturbances and surface-like waves inside macro domain for $y<Y$. If Eq.(8.3) has four real roots then due to Vieta theorem two roots should be negative. Let $s_1<0$ and $s_2>0$ then for $\lambda_1 s_1 + \lambda_2 s_2 > 0$ ; $\lambda_1 + \lambda_2 = 1$ potentials take form:

$$\varphi = \psi = cos(kx - \omega t)[\lambda_1 exp(s_1(y - Y) + \lambda_2 exp(s_2(y - Y))]$$

$$\lambda_1 s_1 + \lambda_2 s_2 = \frac{\omega^2 I_0}{g_y P_0} > 0$$

Taking into account Eq. (5.5; 5.6) Investment and Profits can be presented as:

$$I(t,x,y) = I_0 + \frac{\omega}{d} sin(kx - \omega t) [\lambda_1 exp(s_1(y - Y)) + \lambda_2 exp(s_2(y - Y))] + \frac{p_x}{d}(x - X) + \frac{p_y}{d}(y - Y)$$

$$P(t,x,y) = P_0 + \frac{\omega}{b} sin(kx - \omega t) [\lambda_1 exp(s_1(y - Y)) + \lambda_2 exp(s_2(y - Y))] + \frac{g_x}{b}(x - X) + \frac{g_y}{b}(y - Y)$$

For $s_1<0$ Investment and Profits disturbances for $y<Y$ are proportional to:

$$\sim \lambda_1 sin(kx - \omega t) exp(s_1(y - Y))$$

and they grow up as exponent for $(y-Y)<0$. Let present one else example of possible solution of Eq.(8.1). Let take roots as Eq.(8.6) and assume that one real root $s>0$ and complex roots $s_{3,4}= r+/-i\theta$ have negative real part $r<0$. Then potentials may have form:

$$\varphi = \psi = cos(kx - \omega t)\{\lambda_1 exp(s(y - Y)) + exp(r(y - Y))[\lambda_2 cos(\theta(y - Y)) + \lambda_3 sin(\theta(y - Y)+]$$

$$\lambda_1 + \lambda_2 = 1 \; ; \; s > 0 \; ; \; r < 0 \; ; \; \lambda_1 s + \lambda_2 r + \lambda_3 \theta > 0$$

If real part of $r<0$ of complex roots is negative then Investment and Profits disturbances

$$\sim sin(kx - \omega t) exp(r(y - Y)) [\lambda_2 cos(\theta(y - Y)) + \lambda_3 sin(\theta(y - Y))]$$

and grow up as exponent for $y<Y$. It is interesting that disturbances of Investment and Profits are modulated by factors

$$\lambda_2 cos(\theta(y - Y)) + \lambda_3 sin(\theta(y - Y))$$

and these modulations may establish internal structure of Investment and Profits perturbations far from macro border $y=Y$. Thus Eq.(8.1) admits solutions that describe exponential amplification of Investment and Profits inside macro domain that are induced by small perturbations on border $y=Y$ and such amplifications can impact on macro-finance sustainability. Macro financial surface-like waves for simple model of interaction between Investment and Profits shows hidden complexity of internal financial processes. Description of propagation and interaction of similar surface-like waves for different financial variables might be important for modeling macro-finance sustainability and evolution and these problems require further investigations.

## 5. Conclusions

Description of economic agents by their risk ratings as coordinates on economic space gives ground for new look at macro-finance modeling. Treatment of macro-finance as ensemble of economic agents those fill certain domain on economic space allows describe macro-finance state and evolution via hydrodynamic-like approximations. The most interesting issues concern description of generation, propagation and interactions of various macro-finance waves and their impact on macro financial evolution and sustainability. We believe that waves are necessary element of internal dynamics of any complex system. Macro-finance for sure is one of most complex systems and it's correct description and management requires corresponding complex methods and models. Usage of economic space permits complement common description of macro financial variables as functions of time with description of macro financial variables as functions of time *and coordinates*. That extends methods and techniques for description of macro-finance and opens way for development of macro financial wave models. Macro financial waves that have certain parallels to acoustic waves are described in [5-9]. This paper presents macro financial waves those have similarities with surface waves in fluids. We study a model of simple interaction between Investment and Profits on *2*-dimensional economic space that describe action of two risks. To model common financial relationship: "*more risks – more Profits*" we introduce potentials $H$ and $G$ of "*financial accelerations*" $\boldsymbol{h}=(h_x,h_y)$ and $\boldsymbol{g}=(g_x, g_y)$ that describes linear steady state distributions of Investment and Profits on economic space. We underline distinctions between equilibrium states in the meaning of statistical physics and steady state distributions on economic space in macro-finance. We consider macro-finance as strongly "non-equilibrium" system under transitions from one steady state to another. Description of steady states could allow study dependence of steady distributions of macro financial variables on risk coordinates and model financial policy that could manage transitions from one macro financial steady state to another. Economic agents of macro-finance system fill domain with borders defined by minimum and maximum risk grades. Disturbances of macro financial Investment and Profits near risk border of macro domain may induce waves that propagate along risk border *alike to* surface waves in fluids. Investment and Profits surface-like risk waves transfer disturbances from one part of macro-finance domain to another and carry out interactions between different macro financial variables. Even simplest examples of surface-like waves models admit exponential amplifications of disturbances of macro financial variables like

Investment or Profits inside macro domain. Thus small perturbations of Profits at risk border *y=Y* can cause huge shocks of Profits in relatively secure area for *y<Y* and become origin of macro financial instabilities.

In this paper we study macro-finance model for *local approximation* that proposes that financial transactions are performed between agents with coordinates near same point *x*. In other words, we assume that agents undertake financial transactions with agents with same risk ratings only. Such simplification allows describe interaction between economic variables by Eq. (3.1; 3.2; 5.1-5.4) by simple differential operators. Even simplified model demonstrate extreme complexity and uncovers variety of macro financial waves. Macro-finance model with "action-at-a-distance" financial transactions between agents described in [8].

Description of macro-finance on economic space arises many difficult problems. Main problems concern econometrics. At present there are no econometric data sufficient for modeling on economic space. To do that one should develop unified risk assessment methodologies that can map risk ratings of any economic agents on $R^n$. Risks benchmarking can establish econometric validation procedures and may develop econometric measurements with accuracy alike to measurements in physics. Risk assessment measurements and comparisons can permit verify initial model assumptions and validate or correct macro-finance models, compare predictions with observed data and outlines causes of disagreements between predictions and observations. That requires cooperation of Central Banks and Financial Regulators, Rating Agencies and Market Authorities, Businesses and Government Statistical Bureaus, Academic and Business Researchers, etc. We believe that these efforts will help better understand and manage macro-finance.